\documentclass[a4paper,12pt]{article}

\usepackage[T1]{fontenc}
\usepackage{amssymb,empheq,amsmath} 
\usepackage{physics} 
\usepackage{algorithm} 
\usepackage[noend]{algpseudocode} 
\usepackage[caption=false]{subfig} 
\usepackage{appendix} 
\usepackage{rotating} 

\def\bC{\mathbb{C}}
\def\cD{\mathcal{D}}
\def\cH{\mathcal{H}}
\def\cO{\mathcal{O}}
\def\cP{\mathcal{P}}
\def\cT{\mathcal{T}}
\DeclarePairedDelimiter\ceil{\lceil}{\rceil}

\begin{document}

\title{Practical implementation of a quantum backtracking algorithm}
\author{Simon Martiel$^1$ \and Maxime Remaud$^1$}
\date{%
    $^1$Atos, Quantum R\&D, 78340 Les Clayes-sous-Bois, France\\%
}

\maketitle

\begin{abstract}
In previous work, Montanaro presented a method to obtain quantum speedups for backtracking algorithms, a general meta-algorithm to solve constraint satisfaction problems (CSPs). In this work, we derive a space efficient implementation of this method. Assume that we want to solve a CSP with $m$ constraints on $n$ variables and that the union of the domains in which these variables take their value is of cardinality $d$. Then, we show that the implementation of Montanaro's backtracking algorithm can be done by using $\cO(n\log{d})$ data qubits. We detail an implementation of the predicate associated to the CSP with an additional register of $\cO(\log{m})$ qubits. We explicit our implementation for graph coloring and SAT problems, and present simulation results. Finally, we discuss the impact of the usage of static and dynamic variable ordering heuristics in the quantum setting.

\end{abstract}

%
%

\section{Introduction}

\noindent\emph{Quantum computing.} Quantum computing is one of the most promising emerging computation technology. Theory promises algorithmic speed ups ranging from quadratic, for unstructured problems, up to exponential for some particular key problems. Besides the some problems such as integer factoring and its obvious applications in cryptology, very few applicable large scale algorithms have been derived or studied.  In 2015 especially, Montanaro \cite{Mon15} presented a general method to obtain speedups of backtracking-based algorithms, relying on Belovs' previous work \cite{Bel13} (merged in \cite{BCJ13}). The algorithm uses Quantum Walks, a well developed quantum algorithmic tool intensively studied in the scope of search algorithms \cite{Amb03,Amb07,CCD03,Kem03,MNRS07,San08,Sze04}.

\noindent\emph{Constraint satisfaction problems.} CSPs form a very general class of problems, that encompass a large set of practical problems. The most famous examples are the Boolean satisfiability problem (SAT) \cite{GPFW97} and the graph coloring problem \cite{MT10}. Both have applications in many fields, like scheduling \cite{Lei79} or timetabling \cite{deW85} for graph coloring, and in computer science or artificial intelligence, among others, for SAT (refer to \cite{GPFW97} for a wide list of applications of SAT).


CSPs have been widely studied and a well known tool taking advantage of their structure to solve them is backtracking. For example, the DPLL \cite{DLL62,DP60} backtracking-based algorithm has been introduced in 1962 and is currently the procedure at the basis of some of the most efficient SAT solvers \cite{ES04,GKSS08,vBe06}. Since backtracking algorithms explore a tree whose vertices are partial solutions to the associated CSP \cite{vBe06}, one can think of using a quantum walk to obtain a speedup.

\noindent\emph{Quantum backtracking.} In 2017, Ambainis and Kokainis \cite{AK17} have dealt with Montanaro's algorithm in depth and in 2018, Aono, Nguyen and Shen \cite{ANS18} used these works to speed up the two most efficient forms of enumeration (a lattice algorithm) known. They claim that it affects the security estimates of several lattice-based submissions to the NIST post-quantum standardization process \cite{CJL16}. In turn, Montanaro \cite{Mon19a} has presented how to get a quantum speedup of branch-and-bound algorithms by using Ambainis-Kokainis' work and his own. Thus, Montanaro's algorithm is of high interest in computing science and cryptography. In 2018, Campbell, Khurana and Montanaro \cite{CKM18} reviewed its complexity when applied to the graph coloring problem and SAT, assuming access to a very large amount of qubits, since they aggressively optimized the circuit depth. The aim of this work is to investigate a memory efficient implementation of this backtracking algorithm, in order to be able to validate the algorithm on small instances via classical emulation.

The paper is organized as follows. Section \ref{Prel} introduces definitions. We discuss the choice of heuristic in section \ref{Heur}. Then, section \ref{Implem} is about the use of the predicate, how to implement it and how to check some generic constraints. We conclude by presenting some of the results we got for graph coloring, thanks to a simulator, in section \ref{Res}.

%
%

\section{Preliminaries} \label{Prel}

Thereafter, we will denote by $\cP$ a constraint satisfaction problem defined by a triple $\langle X,D,C \rangle$, where $X = \{x_1,\dots,x_n\}$ is a set of $n$ variables, $D = [\![1,d]\!]$ is a set of $d$ values and $C$ is a set of $m$ constraints. We will also use $\cD$ to denote the extended domain $D\cup\{\ast\}$ in which each variable can take its value, $\ast$ standing for "the variable has no value".

Let $x$ be an assignment of values to the $n$ variables of a CSP $\cP$. It will be said to be a solution to $\cP$ if it verifies all the constraints in $C$; complete if $\forall x_i \in X,~ x_i \neq \ast$, partial otherwise; valid if it is partial and can be extended to a solution; invalid if it is partial and not a solution.

\subsection{Backtracking algorithm}

A standard approach for solving a CSP $\cP$ is the technique of backtracking. For this, we assume that we have access to a predicate $P$ which can receive an assignment (complete or partial) of values $x$ as argument and returns "true" if $x$ is a solution, "indeterminate" if it is valid, "false" otherwise. We also assume access to a heuristic $h$ which specifies which variable should be instantiated next. The general algorithm is presented as algorithm \ref{Backtracking}.

\begin{algorithm}[ht!]
\caption{General backtracking algorithm }\label{Backtracking}
\begin{algorithmic}[1]
	\Require An assignment of values $x$, access to $h : \cD \rightarrow [\![1,n]\!]$ and to $P : \cD \rightarrow \{true, indeterminate, false\}$.
	\If{$P(x)$ is true} output $x$ and \Return \EndIf
	\If{$P(x)$ is false} \Return \EndIf
	\State $j \leftarrow h(x)$
	\For{$w \in [\![1,d]\!]$}
		\State $y \leftarrow x$ with the $j$-th entry replaced with $w$.
		\State \Return \textbf{Algorithm \ref{Backtracking}}$(y)$
	\EndFor
 \end{algorithmic}
 \end{algorithm}

\renewcommand{\algorithmicloop}{\textbf{Repeat}}
\renewcommand{\algorithmicif}{\textbf{If}}
\renewcommand{\algorithmicelse}{\hspace{.5cm}\textbf{else }}

\subsection{Montanaro's algorithm}

The algorithm presented in \cite{Mon15} is based on a quantum walk on trees. The idea is summarized as follows. Consider a rooted tree $\cT$ with $T$ vertices, labeled $r,1,\dots,T-1$, the vertex $r$ being the root of $\cT$. Hereafter, $A$ (resp. $B$) will denote the set of vertices at an even (resp. odd) distance from $r$ and $x \longrightarrow y$ will mean that $y$ is a child of $x$ in the tree. The quantum walk operates on the space spanned by $\{\ket{x};~x\in \{r\} \cup [\![1,T-1]\!]\}$, and starts in the state $\ket{r}$. It is based on a set of diffusion operators $D_x$, where $D_x$ is the identity if $x$ is a solution, otherwise, diffuses on the subspace spanned by $\{\ket{x}\} \cup \{\ket{y}:x\longrightarrow y\}$. A step of the walk consists in applying the operator $R_BR_A$, where :
\begin{center}
$R_A = \bigoplus_{x\in A} D_x$ and $R_B = \dyad{r} + \bigoplus_{x\in B} D_x$
\end{center}

Thanks to these operators, an algorithm for detecting a solution in a tree can be established (algorithm \ref{Detection}). It is the phase estimation of the operator $R_BR_A$ which allows the quantum walker to go through the paths leading to a solution in $\cT$. By applying the detection algorithm to wisely chosen vertices of $\cT$, it is possible to construct an hybrid algorithm for finding a solution (algorithm \ref{Finding}). Montanaro has shown that it finds a solution in time complexity $\cO (\sqrt{T}n^{3/2}\log{n}\log{1/\delta})$ and if there exists a unique solution, it is shown to be $\cO (\sqrt{Tn}\log^3{n}\log{1/\delta})$. For more details on the quantum walk, refer to \cite{Mon15}.

\begin{algorithm}[ht!]
	\caption{Detecting a solution (algorithm 2 of \cite{Mon15})}\label{Detection}
	\begin{algorithmic}[1]
		\Require Operators $R_A$, $R_B$, a failure probability $\delta$, upper bounds on the depth $n$ and the number of vertices $T$. Let $\beta,\gamma > 0$ be universal constants to be determined.
		\Loop $~K = \ceil{\gamma \log{1/\delta}}$ times:
		\State Apply phase estimation to the operator $R_BR_A$ with precision $\beta/\sqrt{Tn}$.
		\If{the eigenvalue is 1} accept \EndIf
		\EndLoop
		\If{number of acceptances $\ge 3K/8$} \Return "Solution exists" \Else{\Return "No solution"} \EndIf
	\end{algorithmic}
\end{algorithm}

\begin{algorithm}[ht!]
	\caption{Finding a solution (described in subsection 2.1 of \cite{Mon15})}\label{Finding}
	\begin{algorithmic}[1]
		\State Apply algorithm \ref{Detection} to the entire tree.
		\If{it outputs "No solution"} \Return \EndIf
		\State Apply algorithm \ref{Detection} to each child of the root until one outputs "Solution exists".
		\If{this child is a solution} output its label and \Return \Else{go back to step 3 with this child as the root.} \EndIf
	\end{algorithmic}
\end{algorithm}

%
%

\section{Variable ordering heuristics} \label{Heur}

In order to be able to test the algorithm with the means we have, our glance was initially towards the choice of the heuristic. In fact, the largest qubit overhead in the implementation comes from the heuristic implementation, since one has to store variable indexes inside the quantum memory ($\cO(n\log n)$ qubits are needed). Moreover, even though the depth overhead of the heuristic is asymptotically negligible compared to the rest of the algorithm, it seems 
that for instances of reasonable size, this overhead is not negligible.

Since we are interested in optimizing the number of qubits, we chose to deal with static variable ordering (SVO) heuristics, which can be classically precomputed at the start of the meta-algorithm. This has two main benefits: only $\cO(n\log d)$ data qubits are required to store an assignment and we do not have to produce a reversible implementation of a dynamic variable ordering (DVO) heuristic.

\begin{figure}[ht!]
	\centering
	\includegraphics[width=.8\linewidth]{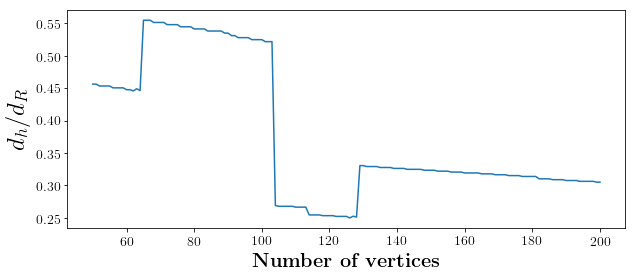}
	\caption{$d_h/d_R$ according to \textit{finalcomplexity.py} for graph coloring given in \cite{Mon19b}.}\label{depth}
\end{figure}

In \cite{CKM18} and \cite{Mon19b}, the implementation of Montanaro's algorithm has been optimized in depth and its complexity has been studied considering the use of a DVO heuristic. However, the benefit of implementing a DVO heuristic is unclear in the range of parameters for which it is claimed in \cite{CKM18} that a graph could be colored in one day (up to approximately 150 vertices). Let $T$ (resp. $T'$) be the number of vertices in the backtracking tree associated with a SVO (resp. DVO) heuristic and $c_T$ (resp. $c_{T'}$) the number of calls to $R_BR_A$ made by Montanaro's algorithm. If we denote the depth of the overhead due to the implementation of a DVO heuristic by $d_h$ and the depth of the operators $R_A$ and $R_B$ without heuristic by $d_R$, we have that using a dynamic heuristic is more efficient than using a static one if: $c_T d_R \ge c_{T'}(d_R + d_h)$, i.e. if $\frac{c_T}{c_{T'}} \ge 1 + \frac{d_h}{d_R}$. This is asymptotically true but thanks to the script computing algorithm \ref{Finding} complexity to solve graph coloring given in \cite{Mon19b}, one can compute an estimate of $\frac{d_h}{d_R}$ (Fig. \ref{depth}). In the considered range of parameters, we can see that using a DVO heuristic would be better than using a SVO one if $c_T \ge \frac{5}{4} c_{T'}$. Unfortunately, as far as we know, no DVO heuristic has been proved to verify such a bound. Thus, the time-saving trick allowed by a DVO heuristic may not be obvious and it might be worth using a SVO heuristic for "small" instances.

The modified algorithm for $R_A$ and $R_B$ is presented as algorithm \ref{Gen_implem}.

\begin{algorithm}[ht!]
	\caption{Implementation of the operator $R_A$}\label{Gen_implem}
	\begin{algorithmic}[1]
		\Require A basis state $\ket{\ell}\ket{v_1}\dots \ket{v_n} \in \bC^{n+1} \otimes (\bC^{d+1})^{\otimes n}$ corresponding to a partial assignment $x_1 = v_1, \dots, x_\ell = v_\ell$. Ancilla registers : $\cH_{\text{anc}}$, $\cH_{\text{children}}$, storing a tuple $(a,S)$, where $a \in \{\ast\} \cup [d]$, 
	$S \subseteq [d]$, initialized to $a=\ast$, $S=\emptyset$.
		\State If $\ell$ is odd, swap $a$ with $v_{\ell}$.
		\State Compute $P(x)$.
	 	\State If $P(x)$ is true, go to step 8.
		\State If $a \neq \ast$, subtract 1 from $\ell$.
	 	\State For each $w \in [d]$, if $P(v_1,\dots,v_\ell,w)$ is not false, set $S = S \cup \{w\}$.
	 	\State If $\ell = 0$, $i=n$, else, $i=1$. Perform $I - 2 \dyad{\phi_{i,S}}$ on $\cH_{\text{anc}}$.
	 	\State Revert steps 5 and 4.
	 	\State Revert steps 2 and 1.
 	\end{algorithmic}
 	$R_B$ is similar, except that: step 1 is preceded by the check "If $\ell = 0$, return"; "odd" is replaced with "even" in step 1; and the check "If $\ell = 0$ is removed from step 6.
 \end{algorithm}
 
In order to test our implementation of Montanaro's algorithm, we used the following SVO heuristics:
\begin{itemize}
	\item Maximum cardinality (MC) \cite{DM89}: chooses the variable with the largest number of neighbors to be the first one and then orders the others by choosing at each step the most connected one with previously ordered ones;
	\item Maximum degree (MD) \cite{DM89}: orders the variables in the decreasing order of the size of their neighborhood;
	\item Minimum width (MW) \cite{Fre82}: orders the variables from last to first by choosing at each step the one having the minimum number of neighbors in the subproblem where previously ordered variables have been deleted.
\end{itemize}

\begin{figure}[ht!]
	\centering
	\includegraphics[width=.9\linewidth]{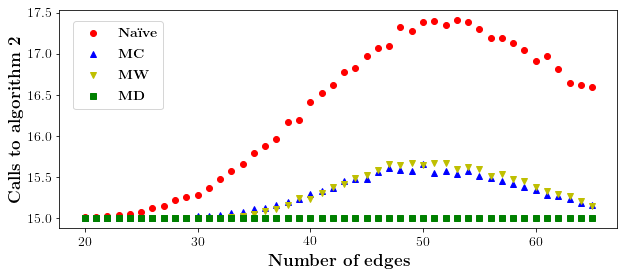}
	\caption{Mean number of calls to algorithm \ref{Detection} depending on the number of edges in a graph with 15 nodes and 4 colors.}\label{heuristic}
\end{figure}

To get an idea of the efficiency of these heuristics compared to the naive one, we emulated Montanaro's algorithm on the graph coloring problem by replacing algorithm \ref{Detection} with a backtracking one in algorithm \ref{Finding}. For fixed values of $n$ and $d$, we generated 1000 random graphs for a varying number of edges, ran algorithm \ref{Finding} on these graphs and counted the number of calls to the emulated algorithm in average. The same pattern arose when $n$ took different values between 10 and 20 and $d$ between 3 and 6 (Fig. \ref{heuristic}). We observed that the MD heuristic offered the best (and optimal) results and that MC and MW heuristics led to very similar results. Of course, we can not extrapolate for higher values of $n$ and $d$.

%
%

\section{Generic implementation} \label{Implem}

Thereafter, we will use the following notations : $\nu = \ceil{\log{(n+1)}}$, $\delta = \ceil{\log{(d+1)}}$ and $\mu = \ceil{\log{(m+1)}}$.

The most straightforward way of implementing the predicate $\cP$ is to compute a logical $AND$ of the result of the evaluation of each constraint over the current variable assignment. This would lead to a circuit using $m+1$ work qubits. We present another solution, a quantum counter, using only $\mu$ qubits.

In the case of the partial predicate, an index $\ell$ stored in a quantum register indicates that the first $\ell +1$ variables have been assigned a value. 
Therefore, if is sufficient to check the constraints that depends on at least one variable in $\{x_{i_1},\dots,x_{i_\ell}\}$. To find out if a constraint has to be checked, we use a system of comparison of values, one of which is quantum and the other classical. The overall process requires $\cO(m)$ additions on $\nu$ qubits.

Similarly, if we apply the detection algorithm to a vertex located at the $\ell$-th level of the backtracking tree, it is unnecessary to check the constraints that depend solely on the variables in $\{x_{i_1},\dots,x_{i_{\ell -1}}\}$.

Thanks to these optimizations, the deeper we go in the backtracking tree, the more qubits, quantum gates and time are saved.

\subsection{How to implement a predicate}

Let $\cP = \langle X,D,C \rangle$ be a CSP. For all $i \in [\![1,n]\!]$, the binary representation of the value assigned to $x_i$ is denoted by $v_i$. The symbol $\ast$ will be encoded by 0.

\begin{figure}[ht!]
	\captionsetup[subfigure]{labelformat=empty}
	\begin{minipage}{.39\linewidth}
		\centering
		\subfloat[(a)]{\label{circ1}\includegraphics[width=\textwidth]{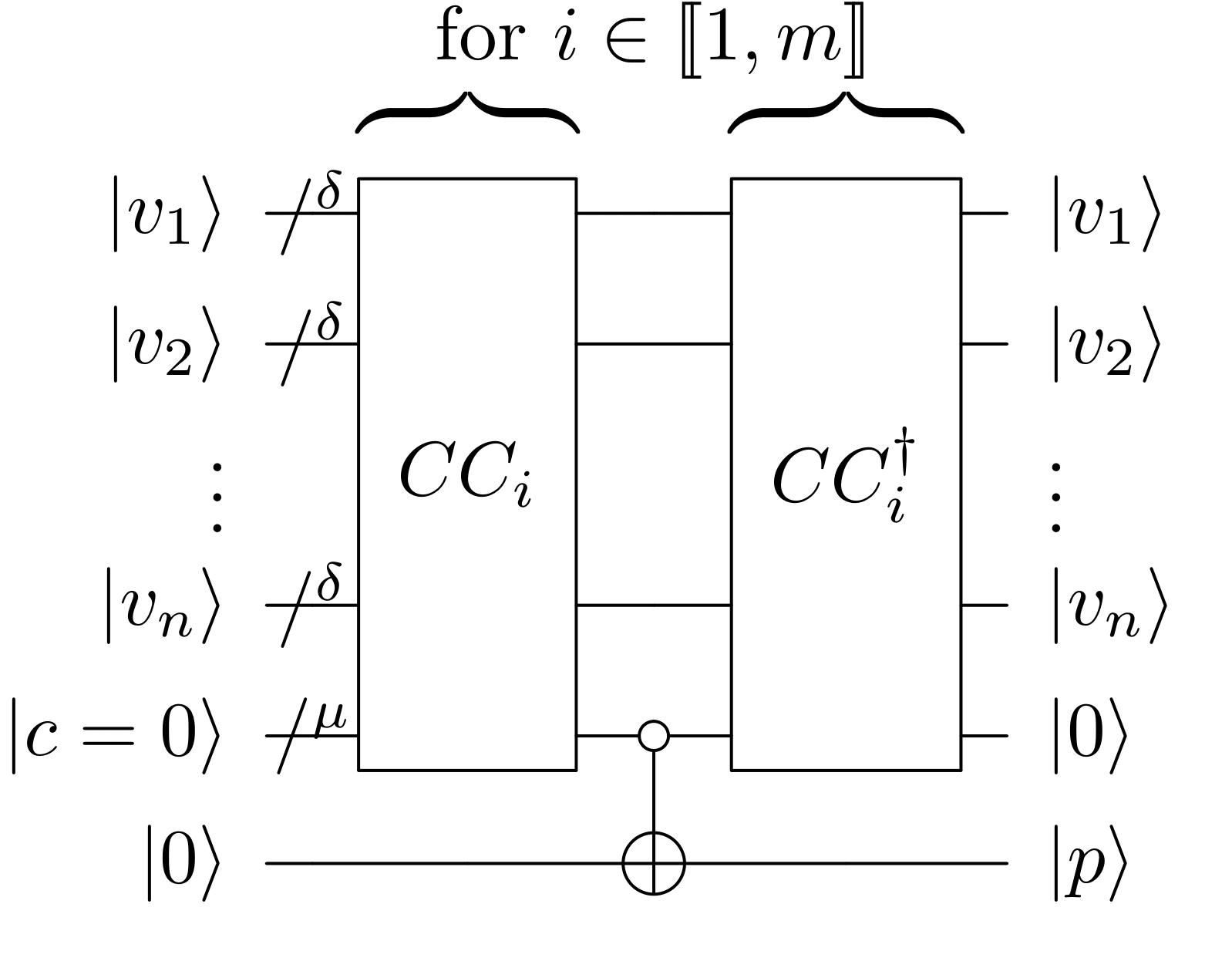}}
	\end{minipage}
	\begin{minipage}{.59\linewidth}
		\centering
		\subfloat[(c)]{\label{circ3}\includegraphics[width=\textwidth]{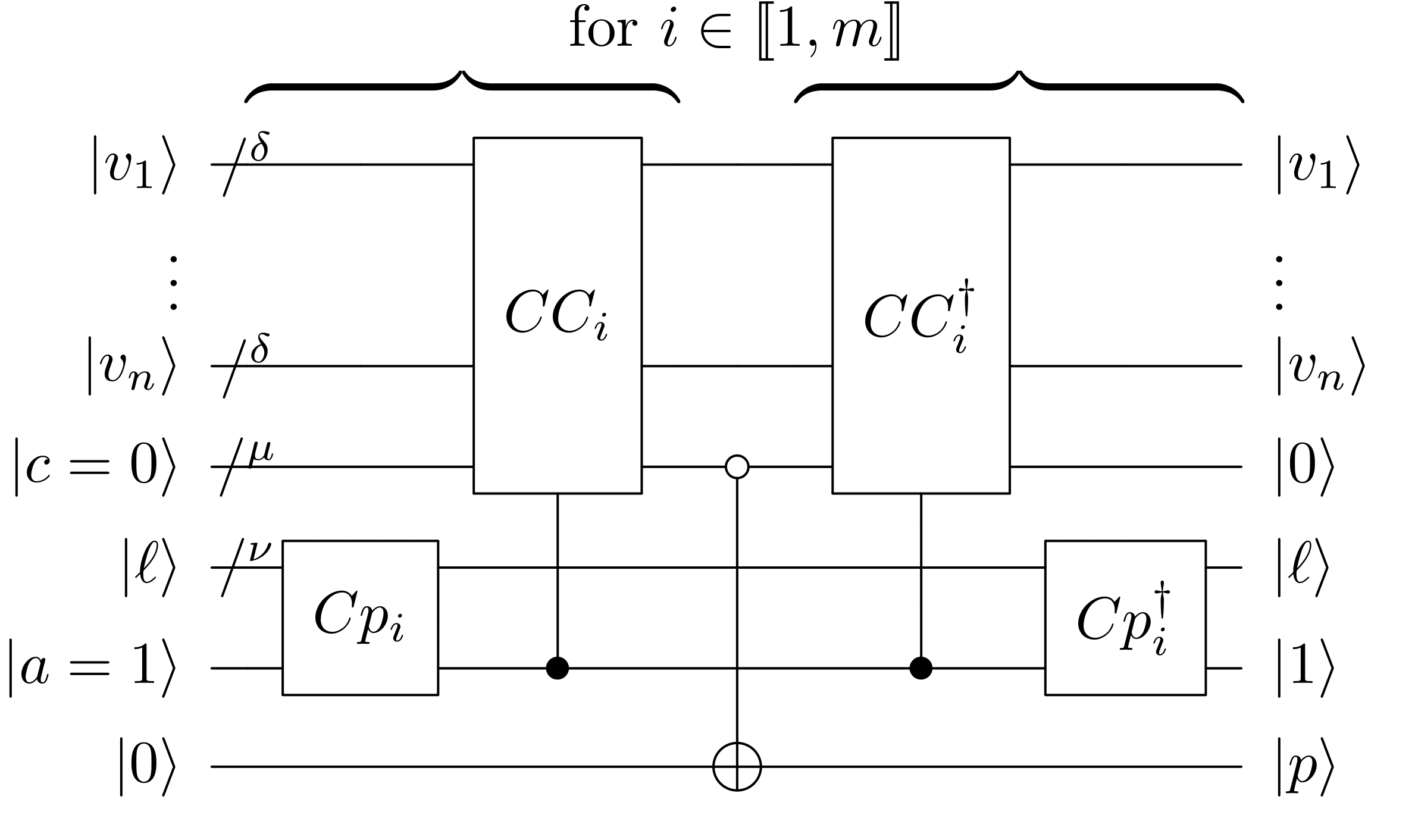}}
	\end{minipage}\par\medskip
	\centering
	\subfloat[(b)]{\label{circ2}\includegraphics[width=.6\textwidth]{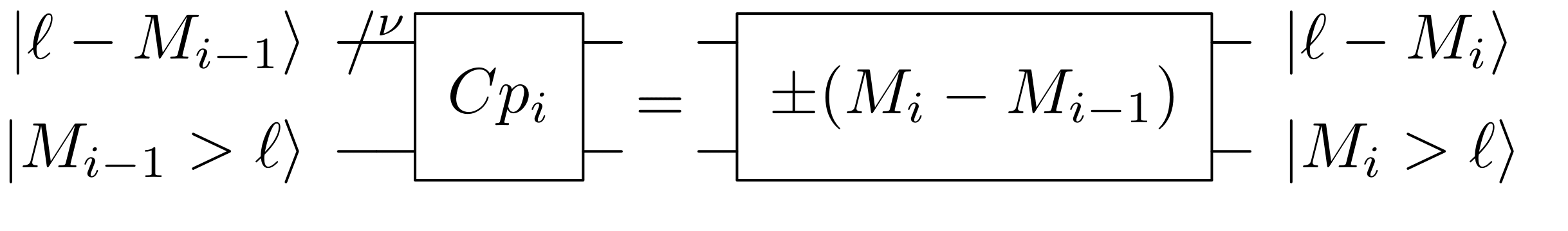}}
	\caption{Circuit (a) computes $P(v_1\cdots v_n)$. If the assignment is a solution, then $p=1$, else $p=0$. Circuit (b) is a comparison operator. The operation is done modulo $2^{\nu +1}$. Circuit (c) computes $P(v_1,\dots,v_{\ell +1})$. If the partial assignment is not invalid, then $p=1$, else $p=0$.}\label{circuits1}
\end{figure}

The circuit \ref{circ1} is used to verify if an assignment is solution to $\cP$, by checking for all $i \in [\![1,m]\!]$ the constraint $C_i \in C$. An ancillary register $c$ is used as a counter. In order to check $C_i$, a subroutine depending on the set $Y_i$ of involved variables in $C_i$ will be used and will increase the counter if $C_i$ is not verified. Once all the constraints have been checked, the counter will be equal to the number of constraints that are violated. Thus, if it is 0, we set an ancillary qubit to 1. Then, we reverse the $C_i$ checking operations to reset the counter to 0. At the end of the circuit, we have in the ancillary qubit the wanted value: $P(v_1,\dots,v_n)$.

We also want to be able to verify if a partial assignment $x_1 = v_1, \dots, x_{\ell +1}= v_{\ell +1}$ is valid for $\cP$. In this aim, we have to check for all $i \in [\![1,m]\!]$ the constraint $C_i$ if $\max_{x_i \in Y_i} \{i\} -1 \le \ell$. In the following, $\forall i \in [\![1,m]\!]$, $M_i$ will denote $\max_{x_i \in Y_i} \{i\} -1$ and $M_0 = 0$.

In order to compute the comparison operator, we add a bit (most significant one) to $\ket{\ell}$ (call it $\ket{a}$, initialize it to $\ket{0}$) and use the following procedure (\ref{circ3}):
\begin{enumerate}
	\item If $M_{i}-M_{i-1}>0$, subtract it from $\ket{\ell -M_{i-1}}$, otherwise, add it to $\ket{\ell -M_{i-1}}$;
	\item If $M_i>\ell$ then some overflow will occur and thus, $\ket{a}$ will be flipped. Use it to control the $C_i$ checking operation.
\end{enumerate}

In circuit \ref{circ2}, the ancillary qubit $\ket{a}$ is set to 1. Thus, the comparison operator will flip $\ket{a}$ if $M_i>\ell$, i.e. if $\max_{x_i \in Y_i} \{i\}>\ell +1$. Thanks to $\ket{a}$ we can control the $C_i$ checking operation. For the rest of the circuit, the idea is the same as the one of the circuit \ref{circ1}.

Within the scope of an optimization of the depth of our implementation, note that constraint checking operations can easily be parallelized (figure \ref{parallel}). We just have to divide $C$ in $k\in\mathbb{N}$ sets of $\frac{m}{k}$ constraints and use a copy of the $v_i$ registers and a counter for checking each one. This involves using $\cO(kn\log n)$ qubits but the depth of the predicate would be divided by $k$ (we just have then to fan-out/fan-in $v_i$ registers to $k$ copies, which can be done in depth $\cO(\log k)$).

\subsection{How to check a constraint}

\begin{figure}[ht!]
	\begin{minipage}{.34\linewidth}
		\centering
		\subfloat[]{\label{circ4}\includegraphics[width=\textwidth]{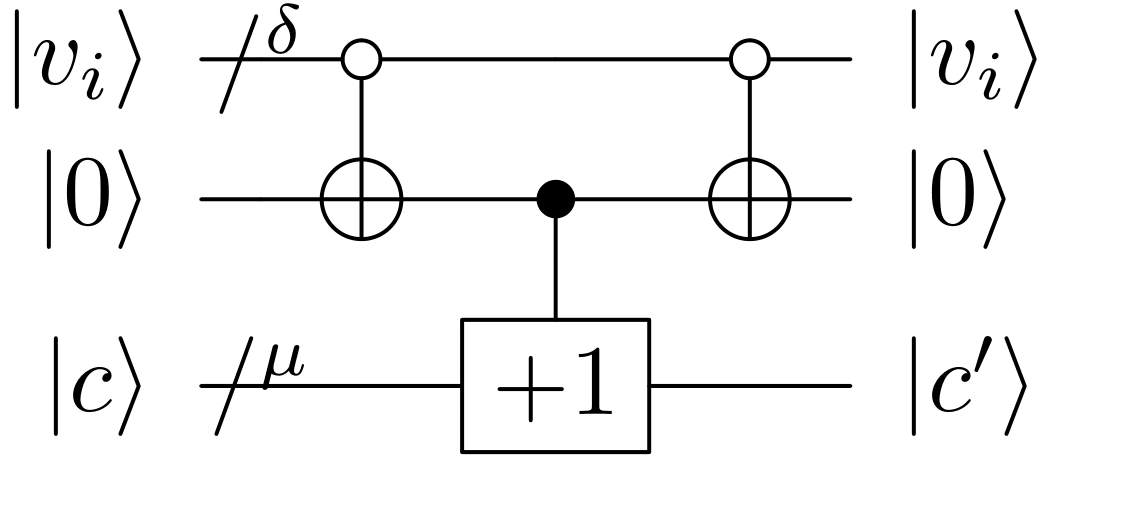}}
	\end{minipage}
	\begin{minipage}{.64\linewidth}
		\subfloat[]{\label{circ5}\includegraphics[width=\textwidth]{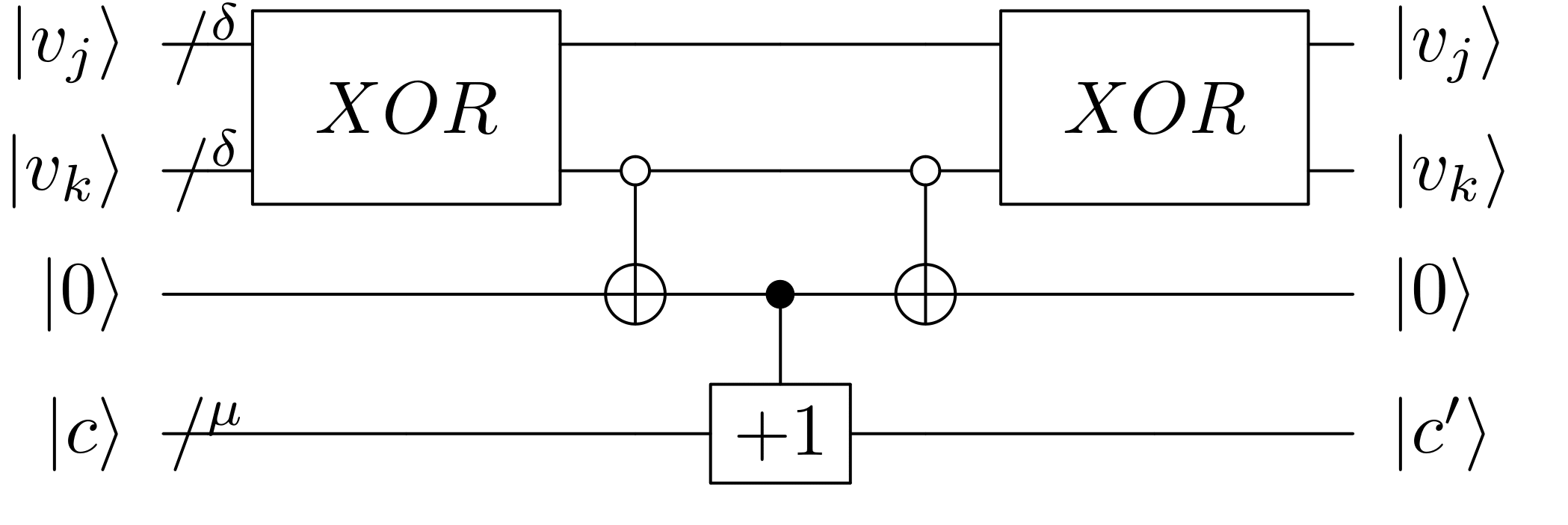}}
	\end{minipage}
	\caption{Circuit (a) checks if the $i$-th variable has been assigned a value. If $v_i =0$, $\ket{c'} = \ket{c+1}$, otherwise $\ket{c'} = \ket{c}$. Circuit (b) checks if the $j$-th and the $k$-th variables have different values. If $v_j = v_k$, $\ket{c'} = \ket{c+1}$, otherwise $\ket{c'} = \ket{c}$.}\label{circuitsgraph}
\end{figure}

The circuit \ref{circ4} is used to check if the variable $x_i$ has been assigned a value (e.g. if the $i$-th vertex of a graph $G$ has been colored). It uses the ancillary register $\ket{c}$ which contains a bit string corresponding to the counter. Since we have chosen that the value 0 means that $v_i=\ast$, we just have to check that $v_i$ is different from 0. The counter is increased by 1 if $x_i$ has not been assigned a value. 

The circuit \ref{circ5} is used to check if the variables $x_j$ and $x_k$ have different values (e.g. if the coloring of the edge between the $j$-th and the $k$-th vertices of a graph is well colored). It uses the ancillary register $\ket{c}$, which is increased by 1 if $v_j \oplus v_k = 0$ (i.e. $v_j = v_k$). For that, we apply a bit-wise XOR to the values of the two variables (the result is stocked in the register of the second value).

For the specific case of boolean variables, we suggest to represent the bit 0 by the quantum state $\ket{10}$, the bit 1 by the quantum state $\ket{11}$ and the unassigned symbol $\ast$ by $\ket{00}$. In the following, we will denote the left qubit by an $L$ in exponent and the right qubit by an $R$ in exponent. Our suggestion stems from the fact that thanks to this choice, it will be simpler to manipulate the variables. The left qubit will allow us to do the negation of the boolean variable without having a side effect on the unassigned values. The right qubit will allow us to distinguish 1 from $\ast$ and 0, which will be useful to check a disjunction of literals.

\begin{figure}[ht!]
	\begin{center}
		\begin{minipage}{.2\linewidth}
			\centering
			\subfloat[]{\label{circ6}\includegraphics[width=\textwidth]{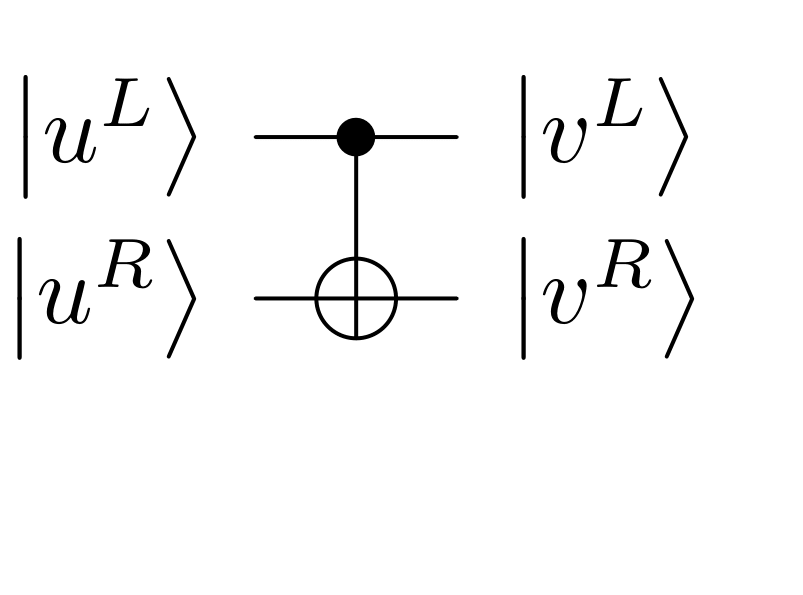}}
		\end{minipage}
		\hspace{2cm}
		\begin{minipage}{.35\linewidth}
			\centering
			\subfloat[]{\label{circ7}\includegraphics[width=\textwidth]{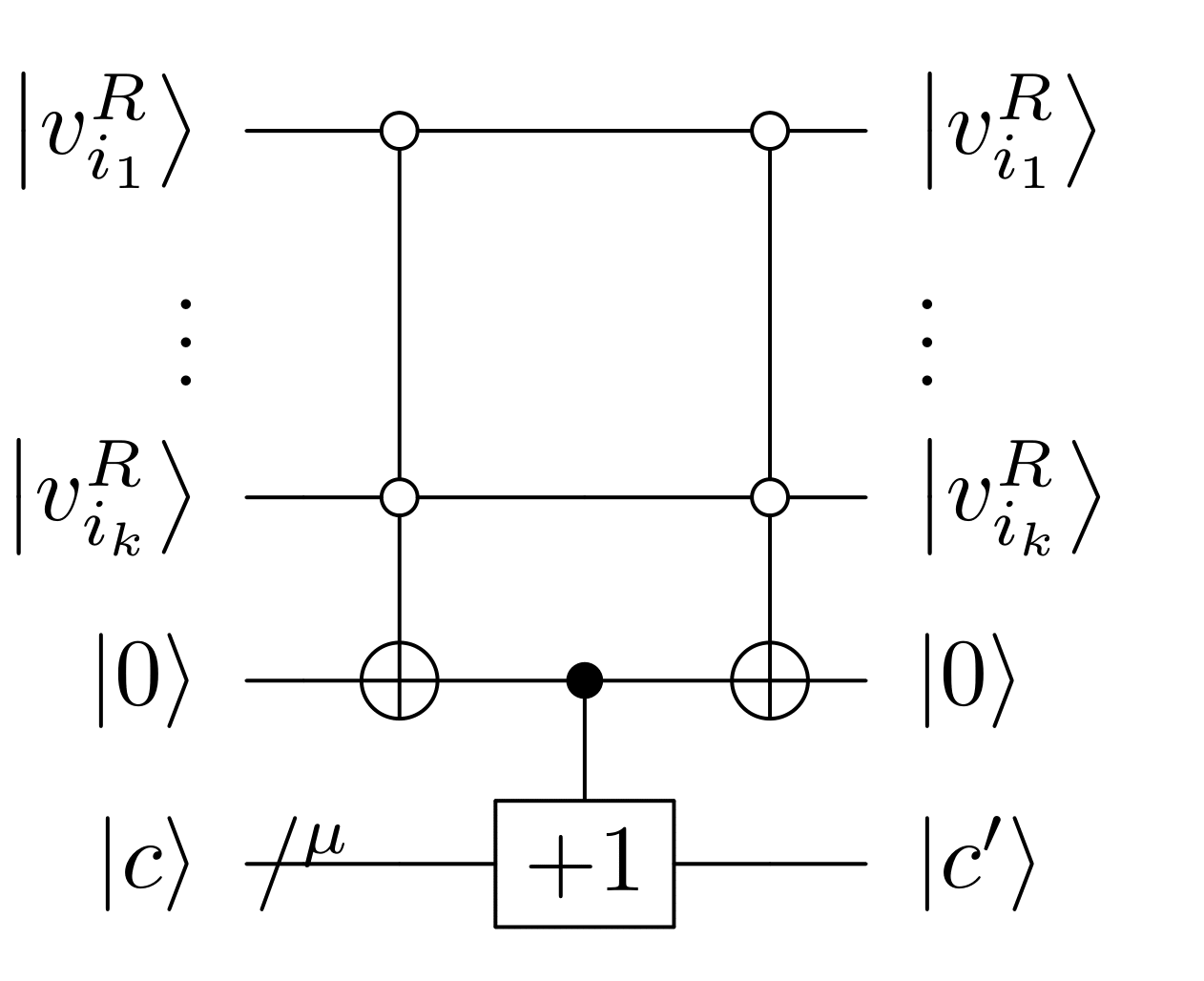}}
		\end{minipage}
	\end{center}
	\vspace{-.5cm}
	\caption{Circuit (a) is the negation of a boolean variable. The value associated with $\ket{v}$ is the negation of the one associated with $\ket{u}$. Circuit (b) checks if the disjunction of the literals $x_{i_1},\dots,x_{i_k}$ is True or False. If $\bigvee_{j \in [\![1,k]\!]}x_{i_j} = 0$, $\ket{c'} = \ket{c+1}$, otherwise $\ket{c'} = \ket{c}$.}\label{circuitssat}
\end{figure}

\subsection{General structure}

Thanks to the precedent subsections, step 2 and most of step 5 of algorithm \ref{Gen_implem} can be realized. The rest of the implementation is most straightforward, circuits of the operators $R_A$ and $R_B$ are given in appendix (resp. Fig. \ref{RA} and \ref{RB}). Note that contrary to what is said in \cite{CKM18}, we just need to add one control qubit to three (resp. one) controlled-Z gates in step 6 to control the whole operator $R_A$ (resp. $R_B$).

%
%

\section{Simulation results} \label{Res}

Now, we present some results of our simulations for the problem of graph coloring. In order to check the constraints, the circuits \ref{circ4} and \ref{circ5} can be used. The colors red (circle), green (square) and blue (triangle) will be denoted by $R$, $G$ and $B$. We will denote any partial assignment $x_1=v_1,\dots,x_\ell=v_\ell$ which is not trivially false by $a^\ell$ and if the values are of importance by $a_{v_1\dots v_\ell}$.

\begin{figure}[ht!]
	\captionsetup[subfigure]{labelformat=empty}
	\begin{minipage}{.24\linewidth}
		\centering
		\subfloat[(a)]{\label{graph1}\includegraphics[width=.6\linewidth]{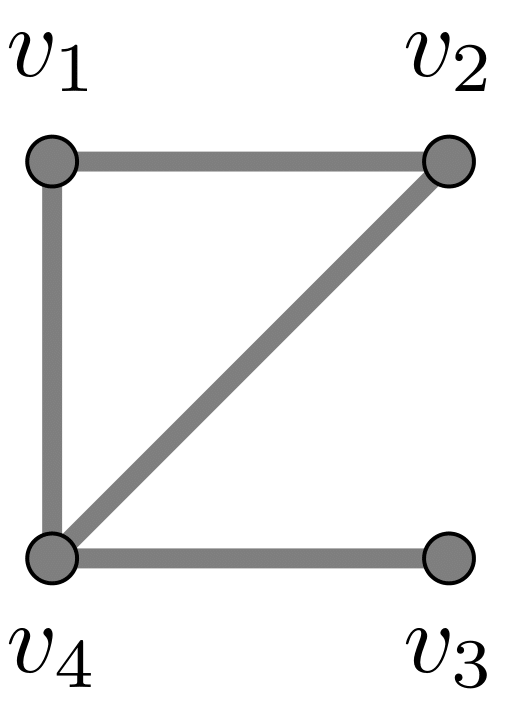}}
	\end{minipage}
	\begin{minipage}{.74\linewidth}
		\centering
		\subfloat[(c)]{\label{table1}\includegraphics[width=.9\linewidth]{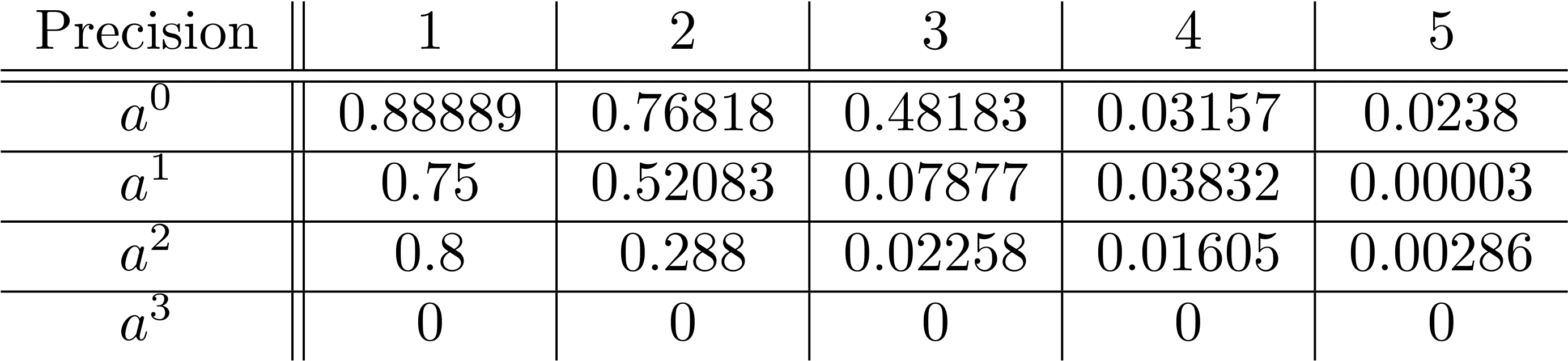}}
	\end{minipage}\par\medskip
	\begin{minipage}{.49\linewidth}
		\centering
		\subfloat[(b)]{\label{tree1}\includegraphics[width=.7\linewidth]{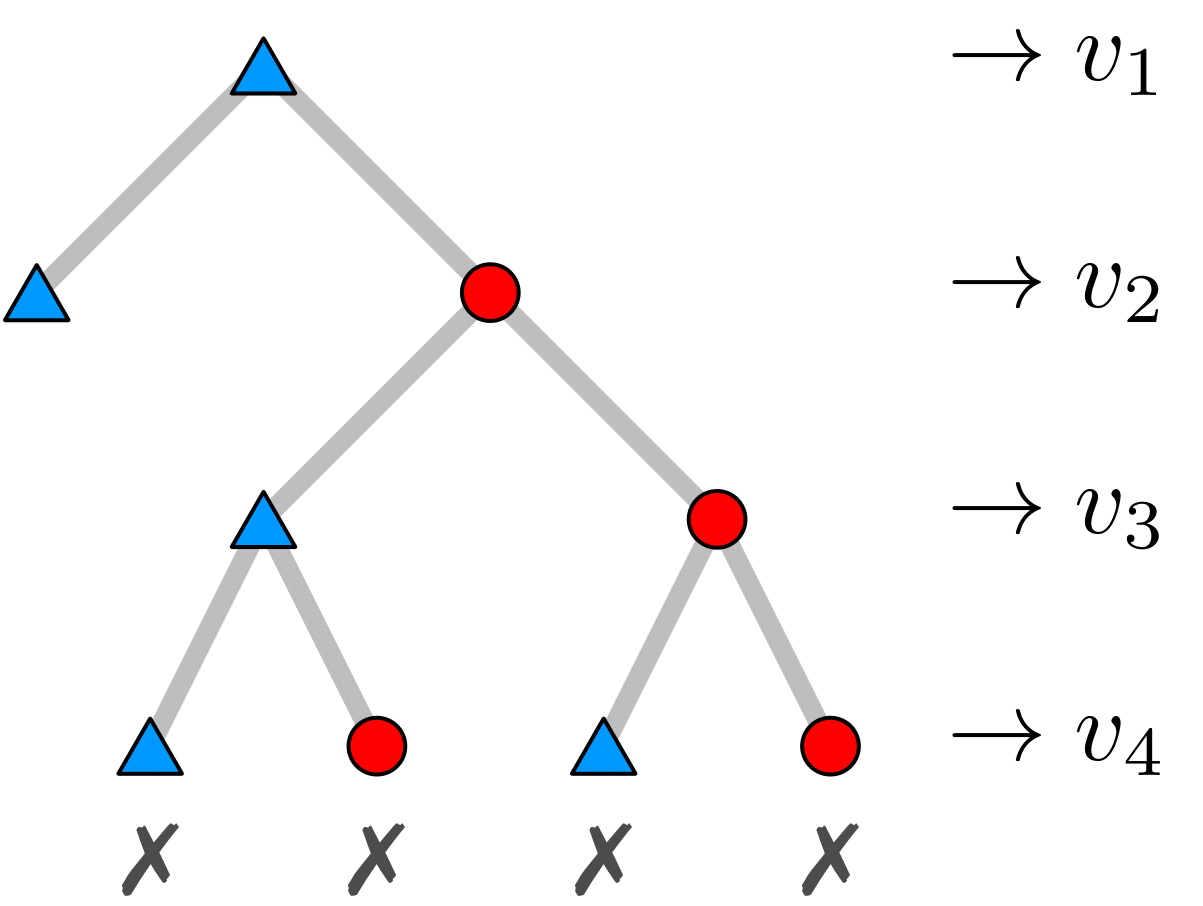}}
	\end{minipage}
	\begin{minipage}{.49\linewidth}
		\centering
		\subfloat[(d)]{\label{graphic1}\includegraphics[width=.6\linewidth]{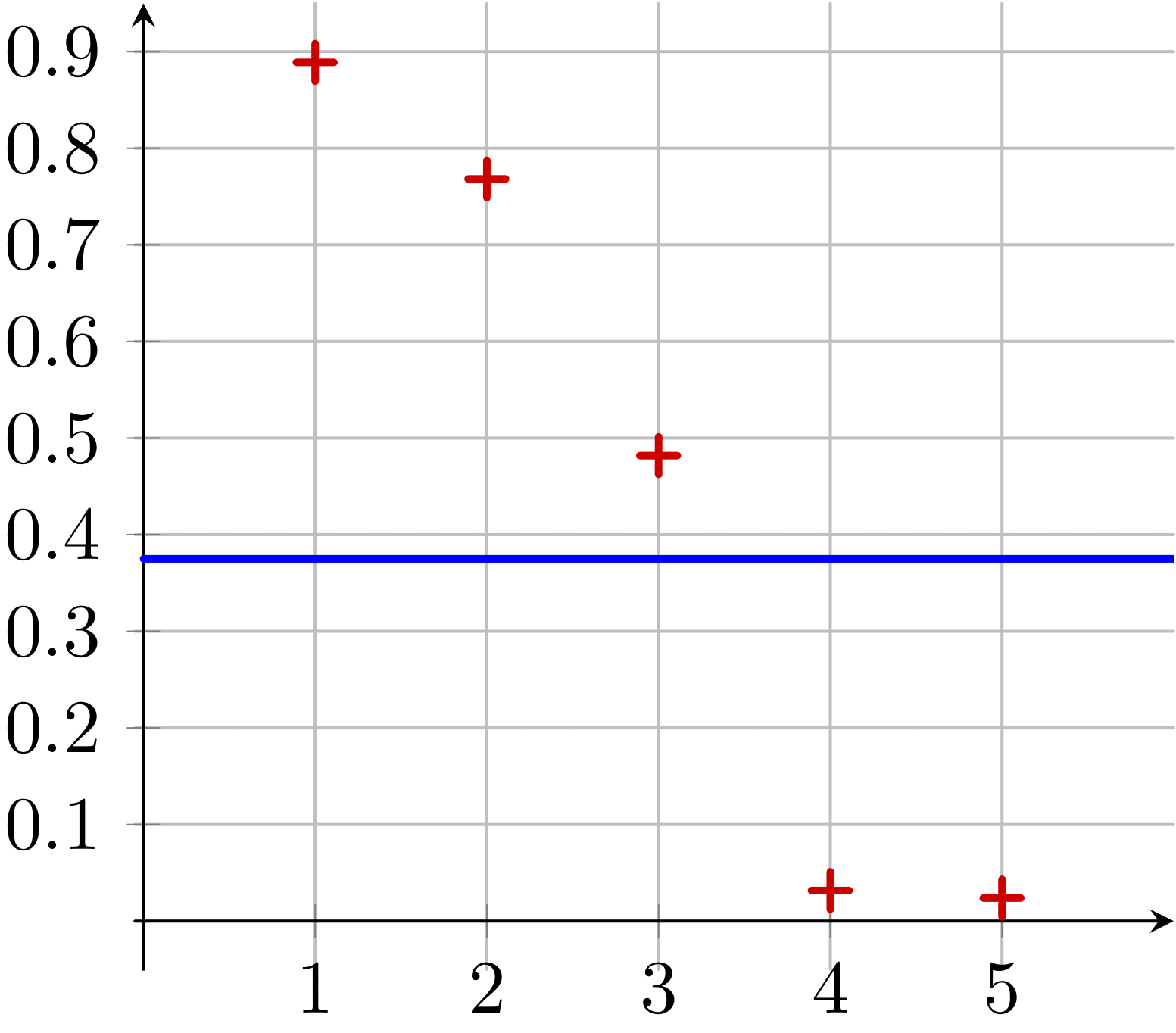}}
	\end{minipage}
	\caption{Fig. (a): a graph. Fig. (b): its associated classical backtracking tree. Tab. (c): results of our simulations. Fig. (d): some of these results as a graph.}\label{example1}
\end{figure}

\begin{figure}[ht!]
	\captionsetup[subfigure]{labelformat=empty}
	\begin{minipage}{.24\linewidth}
		\centering
		\subfloat[(a)]{\label{graph2}\includegraphics[width=.6\linewidth]{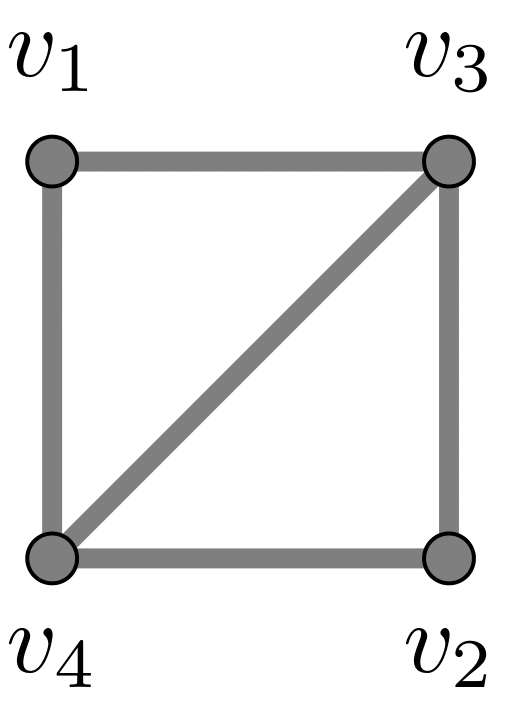}}
	\end{minipage}
	\begin{minipage}{.74\linewidth}
		\centering
		\subfloat[(c)]{\label{table2}\includegraphics[width=.9\linewidth]{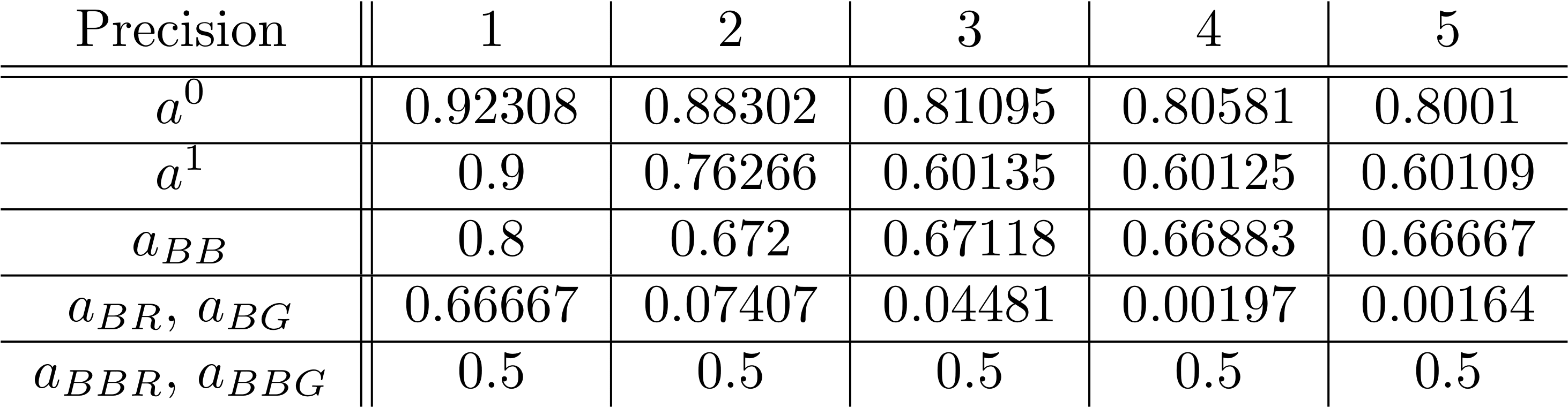}}
	\end{minipage}\par\medskip
	\begin{minipage}{.49\linewidth}
		\centering
		\subfloat[(b)]{\label{tree2}\includegraphics[width=.85\linewidth]{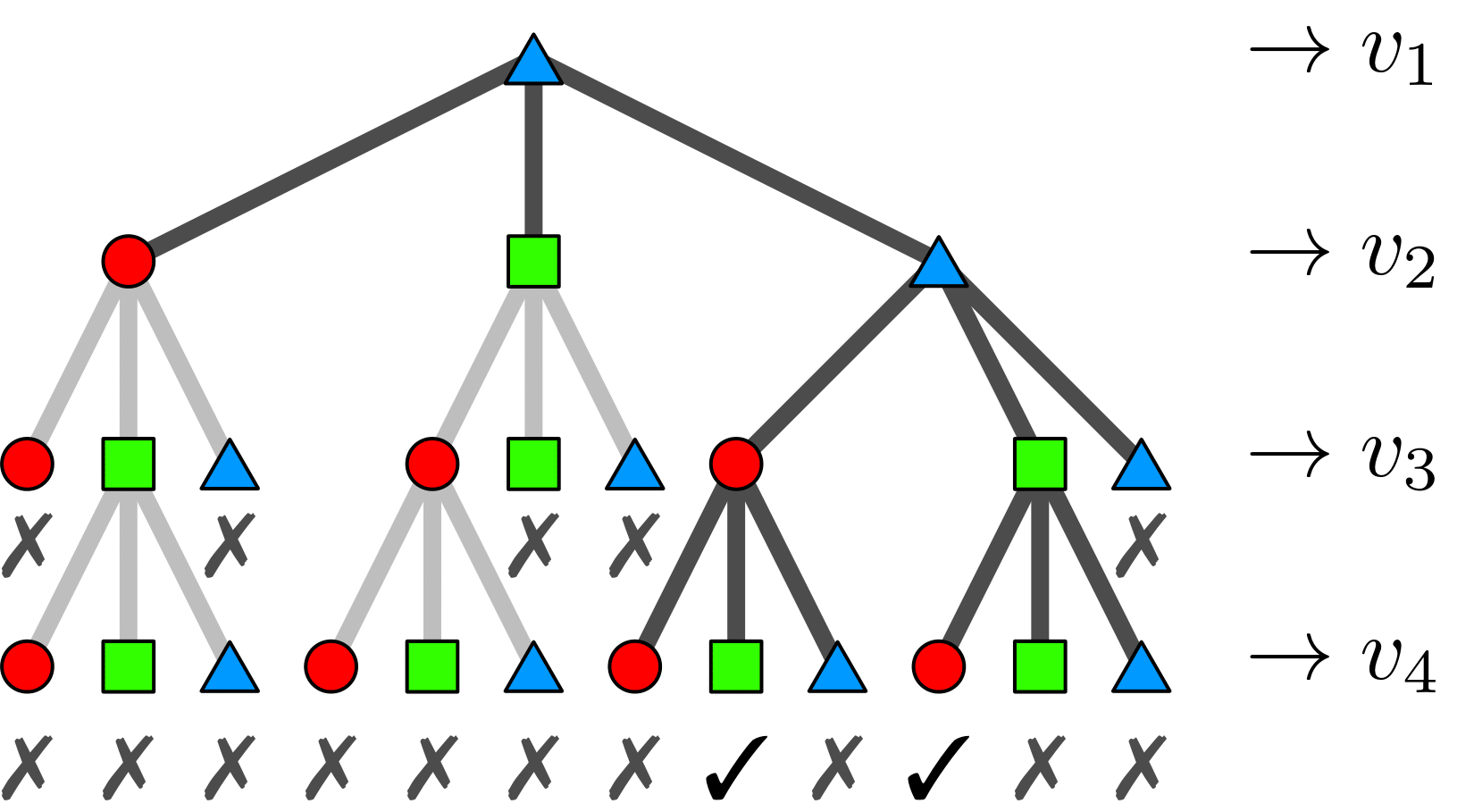}}
	\end{minipage}
	\begin{minipage}{.49\linewidth}
		\centering
		\subfloat[(d)]{\label{graphic2}\includegraphics[width=.6\linewidth]{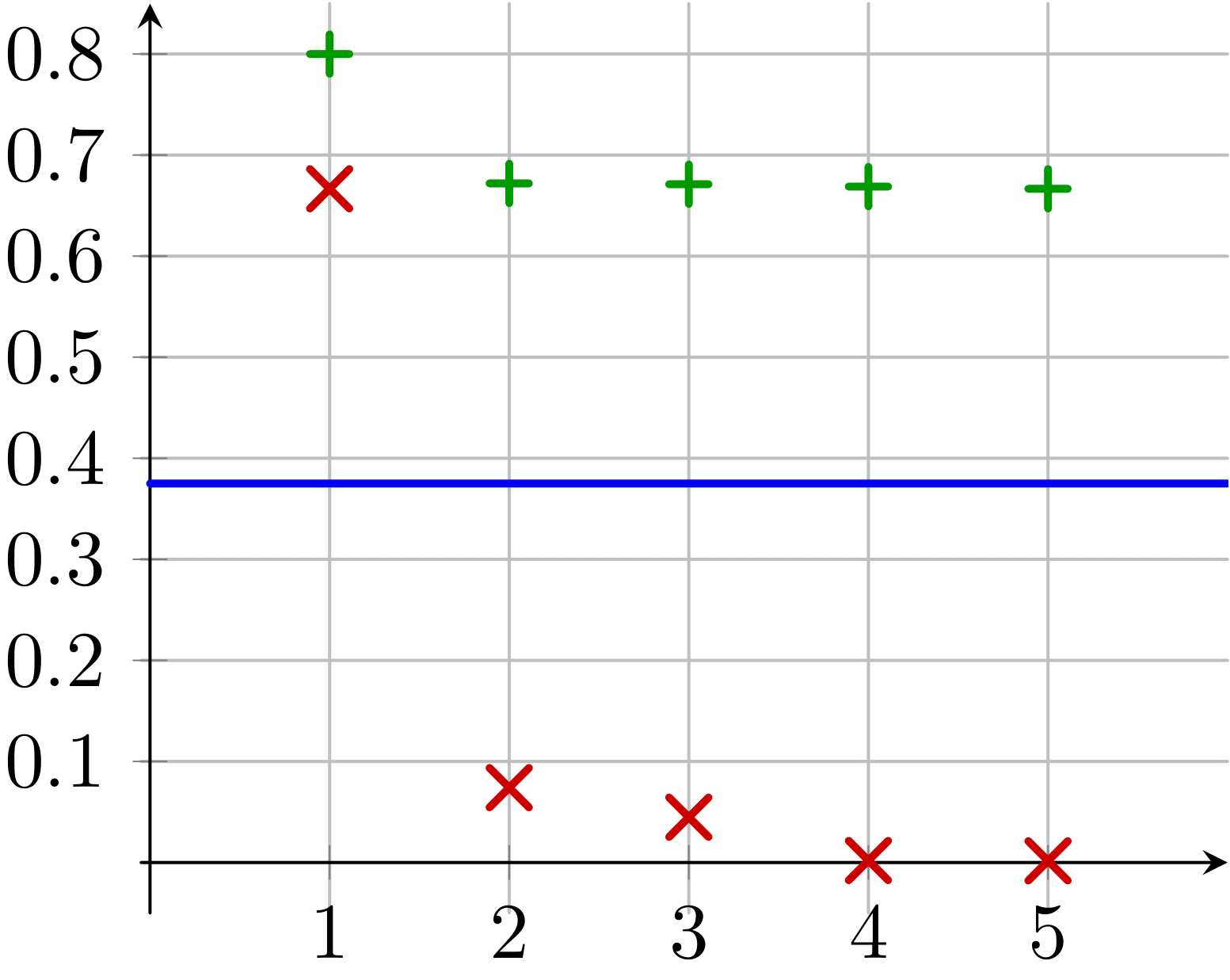}}
	\end{minipage}
	\caption{Fig. (a): a graph. Fig. (b): a subtree of its associated classical backtracking tree ($d=3$). Tab. (c): results of our simulations. Fig. (d): some of these results as a graph.}\label{example2}
\end{figure}

Fig. \ref{table1} (resp. \ref{table2}) shows the classical backtracking tree associated to the graph in Fig. \ref{graph1} (resp. \ref{graph2}), assuming, without loss of generality, that the first vertex has been colored in blue.

By applying the algorithm \ref{Finding} on the root of the complete tree associated to the graph in Fig. \ref{graph1}, it is supposed to detect that there is no possible coloring for this graph with two colors. Thus, the subtree presented here may not be visited thanks to the quantum part. For the second example, the algorithm \ref{Finding} will travel along the dark gray tree (if it outputs all the solutions in Fig. \ref{table2}).

Tab. \ref{tree1} (resp. \ref{tree2}) presents some probabilities of obtaining eigenvalue 1 depending on the precision used for the phase estimation in the algorithm \ref{Detection} and the partial assignment at the root of the subtree on which it is applied. Some of these probabilities (y-axis) depending on the precision (x-axis) are presented in Fig. \ref{graphic1} (resp. \ref{graphic2}).

For the first example, we look at the results where the algorithm has been applied on $a^0$. We can see that for a precision smaller than 4, the algorithm \ref{Detection} will fail with high probability, since the threshold probability of acceptance fixed in \cite{Mon15} is $0.375$ (blue line). For the second example, we look at the results where the algorithm has been applied on $a_{BB}$ (plus signs) and $a_{BR}$ $\&$ $a_{BG}$ (crosses). We can see that for a precision equal to 1, the algorithm \ref{Detection} will fail with high probability.

%
%

\section{Conclusion}

In this paper, we discussed our implementation of Montanaro's algorithm, but an improved quantum algorithm for backtracking has been introduced by Ambainis and Kokainis \cite{AK17}, reducing the queries complexity from $\cO(\sqrt{Tn}\log{\frac{1}{\delta}})$ to $\cO(n^{3/2}\sqrt{T'}\log^2{\frac{n\log{T}}{\delta}})$, where $T'$ is the number of vertices of $\cT$ actually explored by a classical backtracking algorithm. Nevertheless, Montanaro's algorithm can not be left out since it is a component of Ambainis-Kokainis' algorithm.

While Campbell, Khurana and Montanaro \cite{CKM18} assumed access to an extremely large number of physical qubits to propose a depth optimized method to implement Montanaro's algorithm, we have presented techniques minimizing the space usage. For that, we especially looked at the implementation of the predicate and the heuristic. We have proposed the use of a quantum counter for the former and highlighted the fact that up to a certain point, the latter might not be quantumly implemented. However, these propositions are not asymptotically competitive, although our implementation of the predicate could be parallelized to be efficient and could lead to a trade-off between the space usage and the time usage. As far as the heuristic is concerned, it would be interesting to establish a precise resource estimation and define to what extent a SVO heuristic would present benefits compared to a DVO one.

%
%

\bigskip
\noindent\textbf{Acknowledgments.} This work was supported by Atos. The implementation was developed in python using Atos’ pyAQASM library. All simulations were performed on the Atos Quantum Learning Machine. We acknowledge support from the French ANR project ANR-18-CE47-0010 (QUDATA), the QuantERA ERA-NET Cofund in Quantum Technologies implemented within the European Union’s Horizon 2020 Program (QuantAlgo project), and the French ANR project ANR-18-QUAN-0017 (QuantAlgo Project).

%
%

\bibliographystyle{abbrv}

\begin{thebibliography}{10}

\bibitem{Amb03}
A.~{Ambainis}.
\newblock {Quantum walks and their algorithmic applications}.
\newblock {\em International Journal of Quantum Information \textbf{01}(04)
  507--518}, 2003.

\bibitem{Amb07}
A.~{Ambainis}.
\newblock {Quantum walk algorithm for element distinctness}.
\newblock {\em SIAM J. Comput. \textbf{37}(1) 210--239}, 2007.

\bibitem{AK17}
A.~{Ambainis} and M.~{Kokainis}.
\newblock {\em {Quantum algorithm for tree size estimation, with applications
  to backtracking and 2-player games}}.
\newblock {ACM}, 2017.

\bibitem{ANS18}
Y.~{Aono}, P.~Q. {Nguyen}, and Y.~{Shen}.
\newblock {\em Quantum Lattice Enumeration and Tweaking Discrete Pruning}.
\newblock {Springer}, 2018.

\bibitem{Bel13}
A.~{Belovs}.
\newblock Quantum walks and electric networks.
\newblock {\em arXiv:1302.3143}, 2013.

\bibitem{BCJ13}
A.~{Belovs}, A.~M. {Childs}, S.~{Jeffery}, R.~{Kothari}, and F.~{Magniez}.
\newblock {\em Time-efficient quantum walks for 3-distinctness}.
\newblock {Springer}, 2013.

\bibitem{CKM18}
E.~{Campbell}, A.~{Khurana}, and A.~{Montanaro}.
\newblock {Applying quantum algorithms to constraint satisfaction problems}.
\newblock {\em Quantum \textbf{3}(167)}, 2018.

\bibitem{CJL16}
L.~{Chen}, S.~{Jordan}, Y.-K. {Liu}, D.~{Moody}, R.~{Peralta}, R.~{Perlner},
  and D.~{Smith-Tone}.
\newblock {Report on post-quantum cryptography}.
\newblock {\em NISTIR 8105}, 2016.

\bibitem{CCD03}
A.~M. {Childs}, R.~{Cleve}, E.~{Deotto}, E.~{Farhi}, S.~{Gutmann}, and D.~A.
  {Spielman}.
\newblock {\em {Exponential algorithmic speedup by a quantum walk}}.
\newblock {ACM}, 2003.

\bibitem{DLL62}
M.~{Davis}, G.~{Logemann}, and D.~{Loveland}.
\newblock {A machine program for theorem-proving}.
\newblock {\em Communications of the ACM \textbf{5}(7) 394--397}, 1962.

\bibitem{DP60}
M.~{Davis} and H.~{Putnam}.
\newblock {A computing procedure for quantification theory}.
\newblock {\em JACM \textbf{7}(3) 201--215}, 1960.

\bibitem{deW85}
D.~{de Werra}.
\newblock {An introduction to timetabling}.
\newblock {\em European Journal of Operational Research \textbf{19}(2)
  151--162}, 1985.

\bibitem{DM89}
R.~{Dechter} and I.~{Meiri}.
\newblock {\em {Experimental evaluation of preprocessing techniques in
  constraint satisfaction problems}}.
\newblock {Morgan Kaufmann Publishers Inc.}, 1989.

\bibitem{ES04}
N.~{E{\'e}n} and N.~{S{\"o}rensson}.
\newblock {\em {An extensible SAT-solver}}.
\newblock {Springer}, 2004.

\bibitem{Fre82}
E.~C. {Freuder}.
\newblock {A sufficient condition for backtrack-free search}.
\newblock {\em JACM \textbf{29}(1) 24--32}, 1982.

\bibitem{GKSS08}
C.~P. {Gomes}, H.~{Kautz}, A.~{Sabharwal}, and B.~{Selman}.
\newblock {\em {Satisfiability solvers}}.
\newblock {Elsevier}, 2008.

\bibitem{GPFW97}
J.~{Gu}, P.~W. {Purdom}, J.~{Franco}, and B.~W. {Wah}.
\newblock {\em {Algorithms for the satisfiability (SAT) problem: a survey}}.
\newblock {Amer Mathematical Society}, 1997.

\bibitem{Kem03}
J.~{Kempe}.
\newblock {Quantum random walks: an introductory overview}.
\newblock {\em Contemporary Physics \textbf{44}(4) 307--327}, 2003.

\bibitem{Lei79}
F.~T. {Leighton}.
\newblock {A graph coloring algorithm for large scheduling problems}.
\newblock {\em Journal of research of the NBS \textbf{84}(6)}, 1979.

\bibitem{MNRS07}
F.~{Magniez}, A.~{Nayak}, J.~{Roland}, and M.~{Santha}.
\newblock {\em {Search via quantum walk}}.
\newblock Theory of Computing, 2007.

\bibitem{MT10}
E.~{Malaguti} and P.~{Toth}.
\newblock {A survey on vertex coloring problems}.
\newblock {\em International Transactions in Operational Research \textbf{17}},
  2010.

\bibitem{Mon15}
A.~{Montanaro}.
\newblock {Quantum walk speedup of backtracking algorithms}.
\newblock {\em Theory of Computing \textbf{14}(15) 1--24}, 2015.

\bibitem{Mon19b}
A.~{Montanaro}.
\newblock {Data from Quantum algorithms for CSPs}.
\newblock 07 2019.

\bibitem{Mon19a}
A.~{Montanaro}.
\newblock {Quantum speedup of branch-and-bound algorithms}.
\newblock {\em arXiv:1906.10375}, 2019.

\bibitem{San08}
M.~{Santha}.
\newblock {\em {Quantum walk based search algorithms}}.
\newblock Springer, 2008.

\bibitem{Sze04}
M.~{Szegedy}.
\newblock {\em {Quantum speed-up of Markov chain based algorithms}}.
\newblock IEEE, 2004.

\bibitem{vBe06}
P.~{van Beek}.
\newblock {\em {Backtracking search algorithms}}.
\newblock Elsevier, 2006.

\end{thebibliography}

%
%

\section*{Appendix}
\begin{appendix}

\begin{sidewaysfigure}[ht!]
    \includegraphics[width=\textwidth]{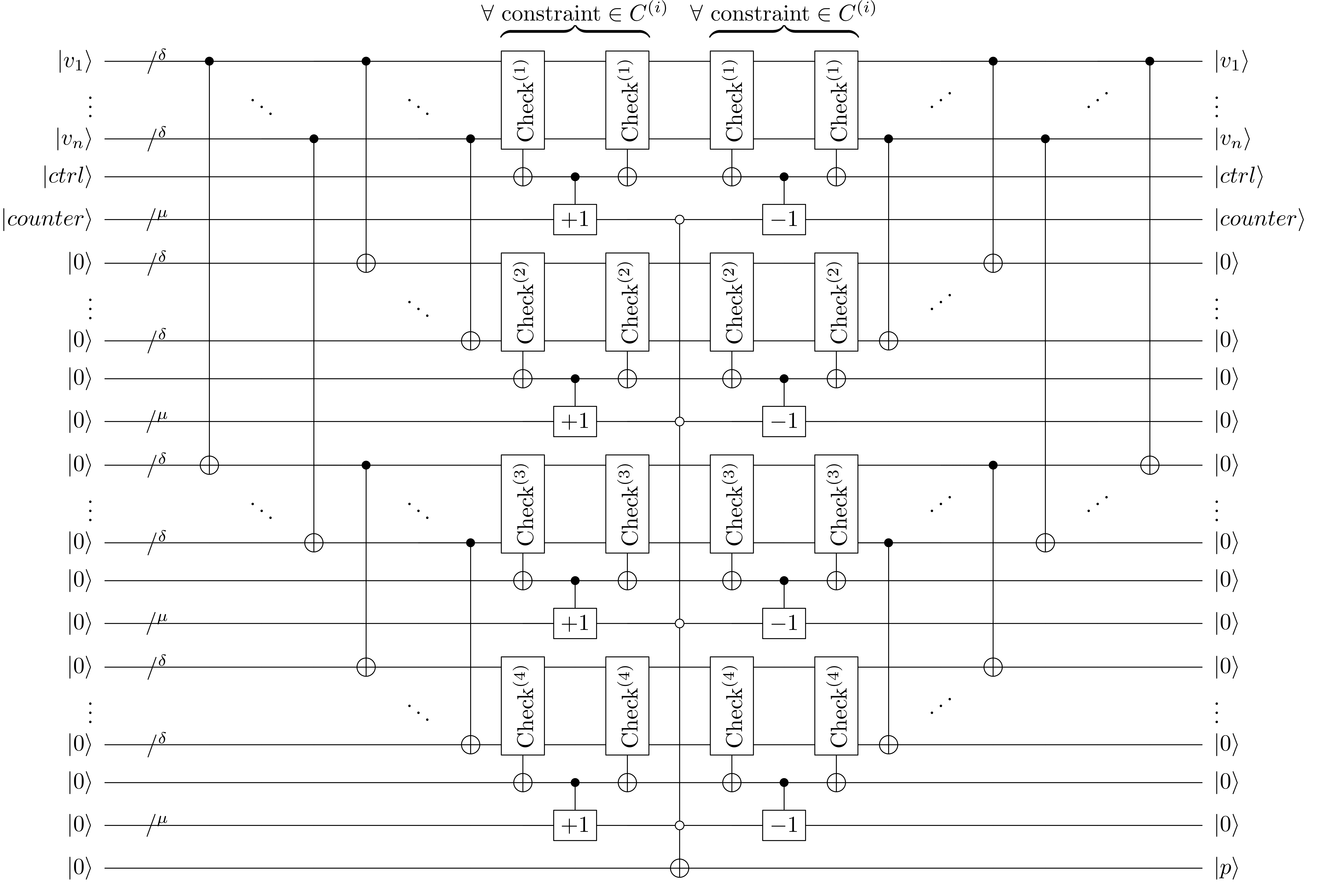}
    \caption{Example of parallelization of the predicate for $k=4$. $\{C^{(i)}, i\in [\![1,4]\!]\}$ is a partition of $C$ and $\abs{C^{(i)}} \le \frac{m}{k} \forall i \in [\![1,4]\!]$. The checking operation of an element in $C^{(i)}$ is denoted by Check$^{(i)}$.}\label{parallel}
\end{sidewaysfigure}

\begin{sidewaysfigure}[ht!]
    \includegraphics[width=\textwidth]{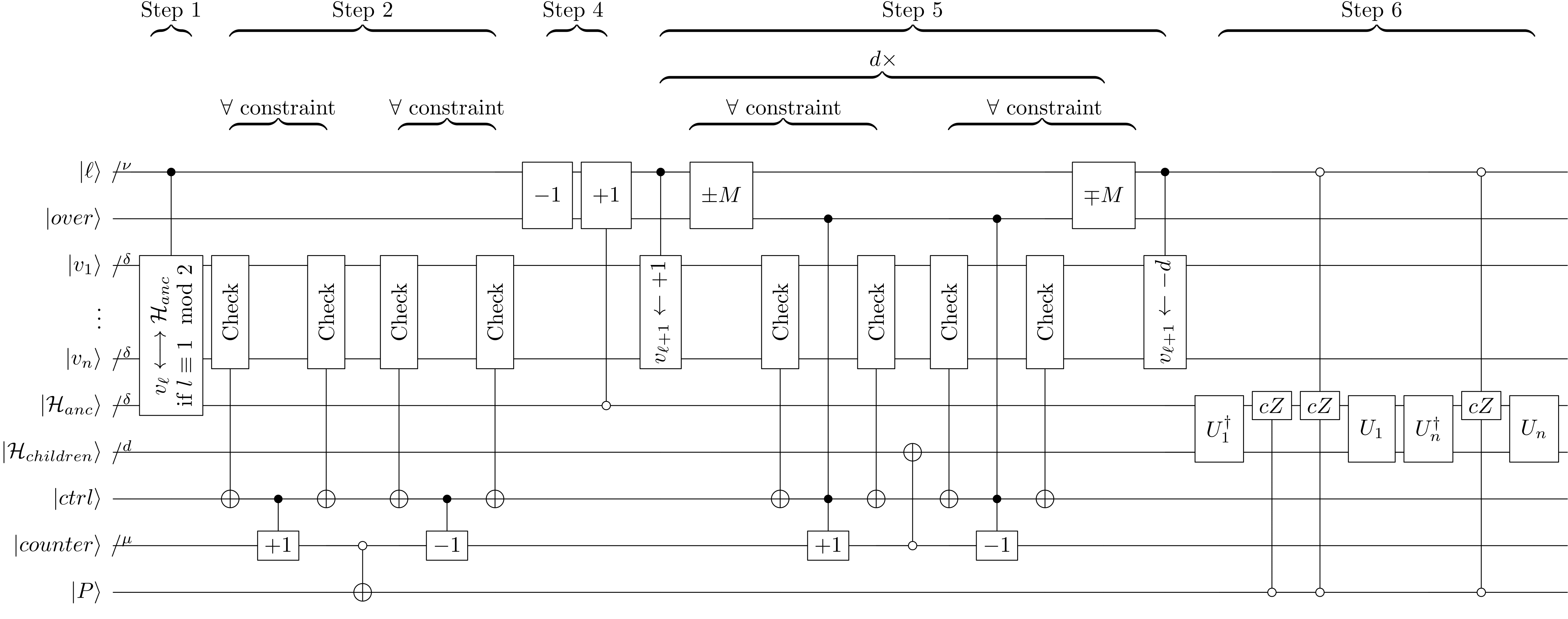}
    \caption{Circuit corresponding to algorithm \ref{Gen_implem} for $R_A$. Steps 7 and 8 are not represented since they consist in reversing steps 5, 4, 2 and 1. Step 3 is realized by controlling cZ (controlled Z) operations in step 6 with $\ket{P}$. In the same way, it is sufficient to control the three cZ in step 6 to control the whole operator, due to the reversibility of the other steps.}\label{RA}
\end{sidewaysfigure}

\begin{sidewaysfigure}[ht!]
    \includegraphics[width=\textwidth]{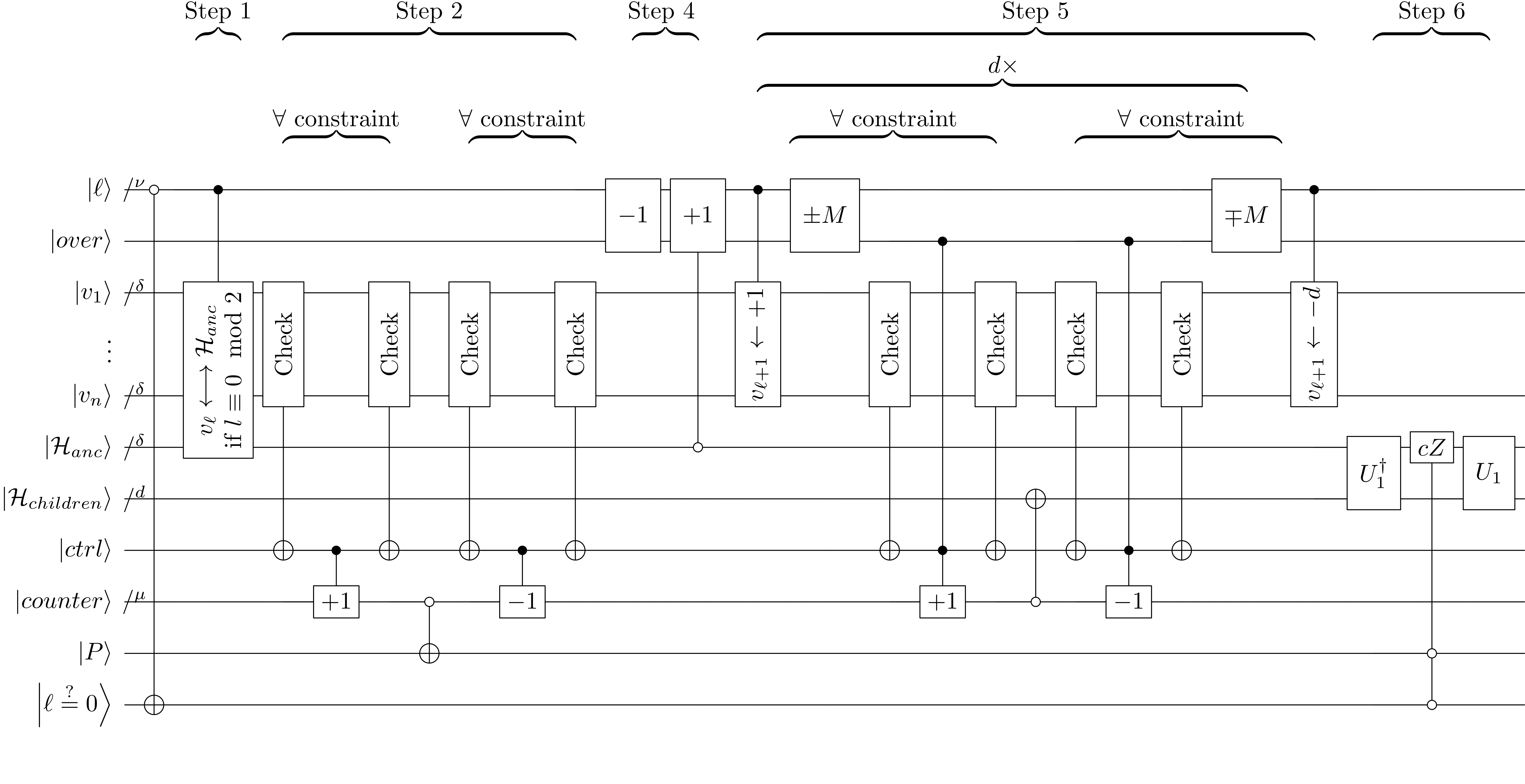}
    \caption{Circuit corresponding to algorithm \ref{Gen_implem} for $R_B$. Steps 7 and 8 are not represented since they consist in reversing steps 5, 4, 2 and 1. Steps "$\ell\stackrel{?}{=}0$" and 3 are realized by controlling cZ (controlled Z) operation in step 6 with $\ket{\ell\stackrel{?}{=}0}$ and $\ket{P}$ respectively. In the same way, it is sufficient to control the cZ operation in step 6 to control the whole operator, due to the reversibility of the other steps.}\label{RB}
\end{sidewaysfigure}

\end{appendix}

\end{document}